\documentclass[aps,prl,reprint,superscriptaddress]{revtex4-1}
\usepackage{natbib}
\usepackage{amsmath}
\usepackage[toc,page]{appendix}
\usepackage{bm}
\usepackage{ amssymb }
\usepackage{graphicx}
\usepackage {epstopdf}
\usepackage{csquotes}

\begin{document}

%\preprint{}

\title{Aharonov-Bohm interference of fractional quantum Hall edge modes}



\author{J. Nakamura}
%\thanks{}
\affiliation{Department of Physics and Astronomy, Purdue University}

\author{S. Fallahi}
%\thanks{}
\affiliation{Department of Physics and Astronomy, Purdue University}
\affiliation{Birck Nanotechnology Center, Purdue University}

\author{H. Sahasrabudhe}
%\thanks{}
\affiliation{Department of Physics and Astronomy, Purdue University}

\author{R. Rahman}
%\thanks{}
\affiliation{School of Electrical and Computer Engineering, Purdue University}

\author{S. Liang}
%\thanks{}
\affiliation{Department of Physics and Astronomy, Purdue University}
\affiliation{Birck Nanotechnology Center, Purdue University}

\author{G. C. Gardner}
%\thanks{}
\affiliation{Birck Nanotechnology Center, Purdue University}
\affiliation{Station Q Purdue, Purdue University}

\author{M. J. Manfra}
\email[]{mmanfra@purdue.edu}
%\thanks{}
\affiliation{Department of Physics and Astronomy, Purdue University}
\affiliation{Birck Nanotechnology Center, Purdue University}
\affiliation{School of Electrical and Computer Engineering, Purdue University}
\affiliation{Station Q Purdue, Purdue University}
\affiliation{School of Materials Engineering, Purdue University}

\date{\today}

\begin{abstract}
We demonstrate operation of a small Fabry-Perot interferometer in which highly coherent Aharonov-Bohm oscillations are observed in the integer and fractional quantum Hall regimes. Using a novel heterostructure design, Coulomb effects are drastically suppressed. Coherency of edge mode interference is characterized by the energy scale for thermal damping, ${T_0=206}mK$ at $\nu=1$. Selective backscattering of edge modes originating in the ${N=0,1,2}$ Landau levels allows for independent determination of inner and outer edge mode velocities. Clear Aharonov-Bohm oscillations are observed at fractional filling factors  $\nu=2/3$ and $\nu=1/3$.  Our device architecture provides a platform for measurement of anyonic braiding statistics.
\end{abstract}

\maketitle

Integer and fractional quantum Hall states are archetypal topological phases of a two-dimensional electron system (2DES) subjected to a strong perpendicular magnetic field \cite{JainBook}. Electronic Fabry-Perot interferometry has been proposed as a means to probe the properties of integer and fractional quantum Hall edge states \cite{Wen1997, DasSarma2005, Halperin2006, Kim2006}; most intriguingly, interferometry may be used to directly observe anyonic braiding statistics \cite{Halperin1984} of fractional quantum Hall quasiparticles. Interference visibility in real devices is limited by finite phase coherence, a particularly acute problem in the fractional quantum Hall regime. Visibility may be improved by decreasing the size of the interferometer so that the path traveled by interfering excitations is shorter. However, attempts to measure interference in small devices have yielded results inconsistent with simple Aharonov-Bohm interference; specifically, the magnetic field oscillation period is found to change with filling factor, and constant phase lines in the gate voltage-magnetic field plane have positive slope rather than the expected negative slope \cite{Zhang2009, Heiblum2010, Goldman2009, Ensslin2013}. This behavior is attributed to Coulomb charging effects \cite{Halperin2007, Halperin2011}, which cause the area of the interferometer to change as the magnetic field is varied. This ``Coulomb-dominated'' behavior masks the Aharonov-Bohm phase and makes braiding statistics unobservable \cite{Halperin2011}. The challenge for measuring robust interference and observing fractional braiding statistics is to create a device small enough to maintain phase coherence, while reducing Coulomb effects so that the device may operate in the Aharonov-Bohm regime.  We report fabrication and operation of an interferometer that overcomes these challenges.

\begin{figure*}[t]
\def\ffile{Structure.pdf}
\centering
\includegraphics[width=\linewidth]{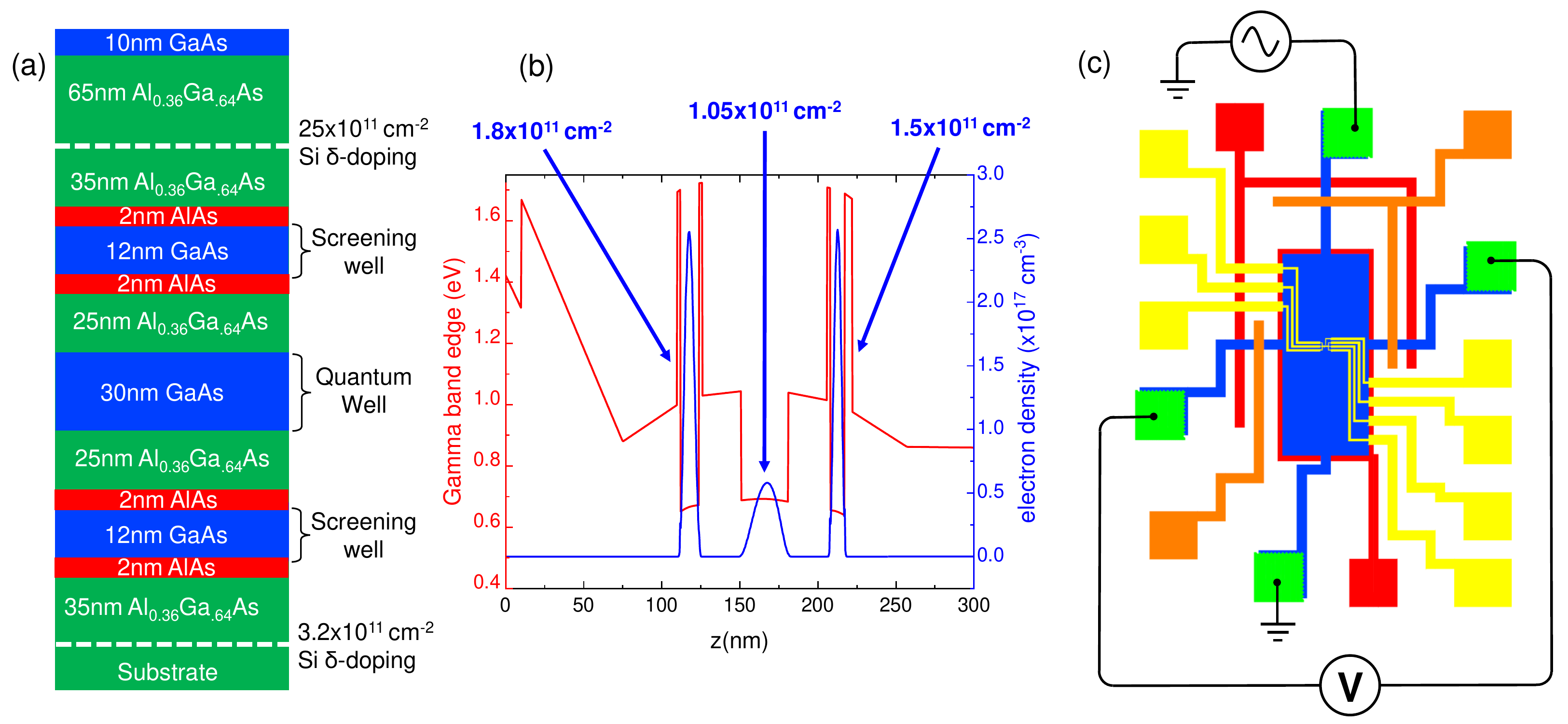}
\caption{\label{Structure}  (a) Layer stack of the GaAs/AlGaAs heterostructure along the growth direction, showing the positions of the GaAs quantum well and screening wells (blue), AlGaAs spacers (green), and AlAs barriers (red). (b) Conduction band edge (red) and electron density (blue) versus growth direction (z-axis) calculated using a self-consistent Schrodinger-Poisson method. The sheet density in each well is indicated. (c) Schematic showing the layout of the mesa (blue), Ohmic contacts (green), surface gates used to isolate the top screening well from the contacts (orange), and the backgate used to isolate the contacts from the bottom screening well (red). The surface gates used to define the interference path are shown in yellow. Additionally, there is a global backgate underneath the mesa (red). A four-terminal measurement circuit is indicated in which current is injected into the Hallbar and the perpendicular Hall voltage is measured; when the interferometer gates are biased to define the interference path, the measured resistance is referred to as the diagonal resistance, $R_D$.}
\end{figure*}

The GaAs/AlGaAs heterostructure was grown by molecular beam epitaxy \cite{Manfra2014, Gardner2016} and is shown in Fig. \ref{Structure} (a). While typical structures utilize a single GaAs quantum well in which the 2DES resides, our structure contains three GaAs wells: a primary quantum well 30nm wide and two additional 12nm wells located on either side of the primary well separated by 25nm AlGaAs spacers. The 2DES under study is located inside the primary GaAs quantum well, while the ancillary wells screen Coulomb effects so that the interferometer may operate in the Aharonov-Bohm regime rather than in the Coulomb-dominated regime \cite{Halperin2007, Halperin2011}. The structure is modulation doped with silicon above the top screening well and below the bottom screening well. In Fig. \ref{Structure} (b) we show the position of the $\Gamma$-band edge (red) and electron density (blue) calculated by the self-consistent Schrodinger-Poisson method \cite{Harshad2018}; the confinement energy in each screening well is tuned to match the experimentally measured densities. This structure is designed to have significantly higher density in the screening wells than in the primary well in order to facilitate strong screening.

Interferometer operation requires transport measurements through the primary quantum well unobscured by parallel conduction through the screening wells. Our device includes narrow gates on the top surface and on the back side of the chip that partially overlay the arms connecting each Ohmic contact to the mesa; this is shown schematically in Fig. \ref{Structure} (c). The surface gates over the Ohmics are negatively biased  at -0.29V; this bias is sufficient to deplete the electrons from the top screening well without depleting either the primary quantum well or the bottom screening well. Similarly, the back side gate over the Ohmics is biased at -150V in order to deplete the bottom screening well, but not the primary quantum well. This eliminates electrical conduction through both screening wells so that only the primary quantum well is probed in measurements. Because these gates are well separated from the gates that define the mesoscopic interference path, the screening wells are still populated in the interferometer and thus available to screen. This selective depletion technique was pioneered to isolate transport in bulk bilayer systems \cite{Eisenstein1990}.  Here we have demonstrated the technique has utility for mesoscale electronic devices as well.

\begin{figure}
\def\ffile{SEMv2.pdf}
\centering
\includegraphics[width=\linewidth]{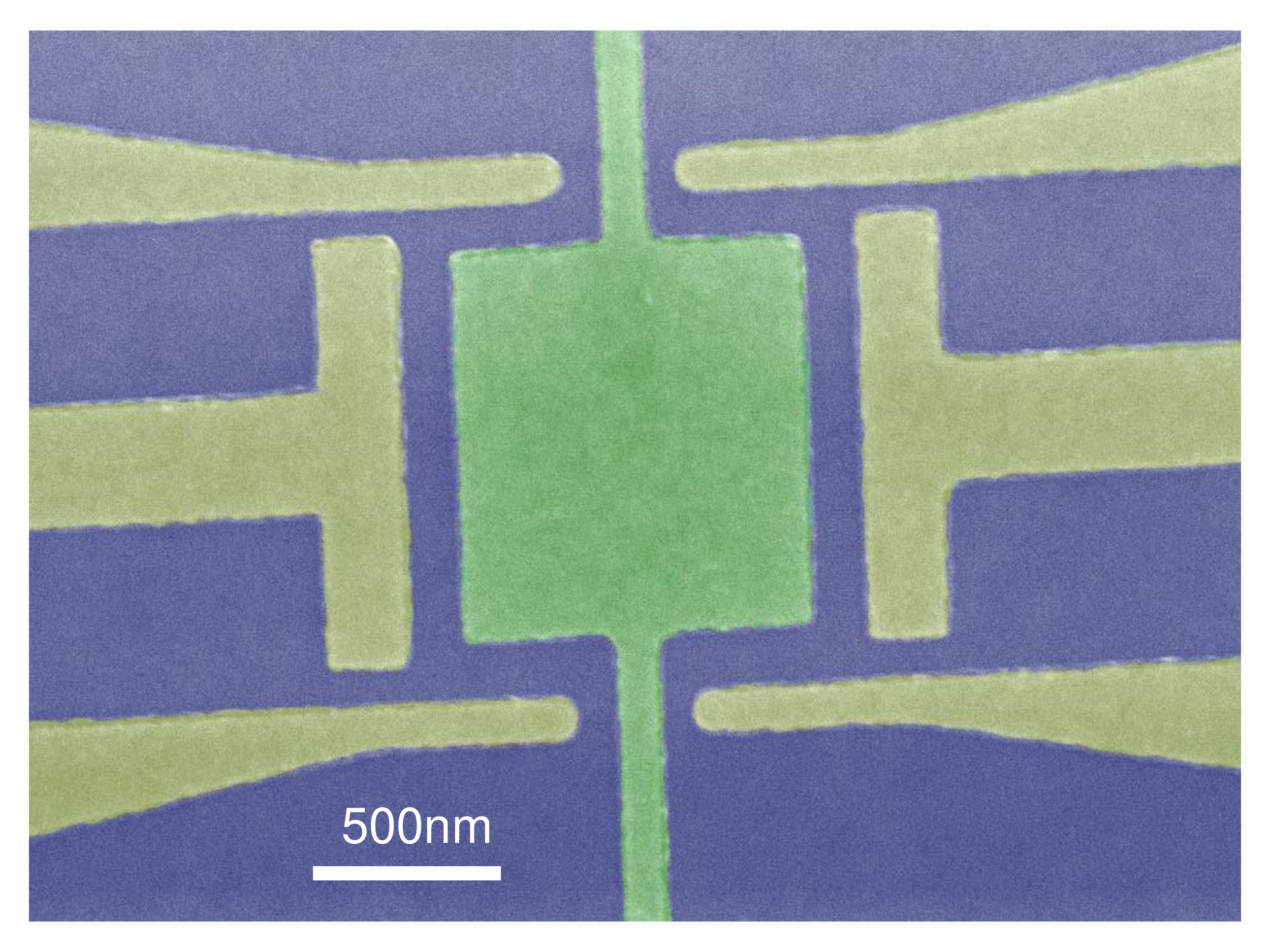}
\caption{\label{SEM}  False color SEM image of the interferometer, located in the center of the Hall bar shown schematically in FIG. 1(c). The device consists of two quantum point contacts to backscatter current and a pair of side gates to define the interference path (yellow); when these gates are negatively biased, the 2DES underneath is depleted, which defines the interference path. An additional gate over the top of the area of the device (green) is grounded for these experiments. }
\end{figure}

In Fig. \ref{SEM} we show a scanning electron microscopy (SEM) image of the interferometer gates. The device consists of two quantum point contacts (QPCs) that form narrow constrictions and a pair of side gates that define the interference path. The gates shown in yellow are negatively biased to deplete electrons from the quantum well and define the interference path; the central top gate (green) is grounded and does not alter the 2DEG density.

\begin{figure}
\def\ffile{Bulk_Blockadev2.pdf}
\centering
\includegraphics[width=\linewidth]{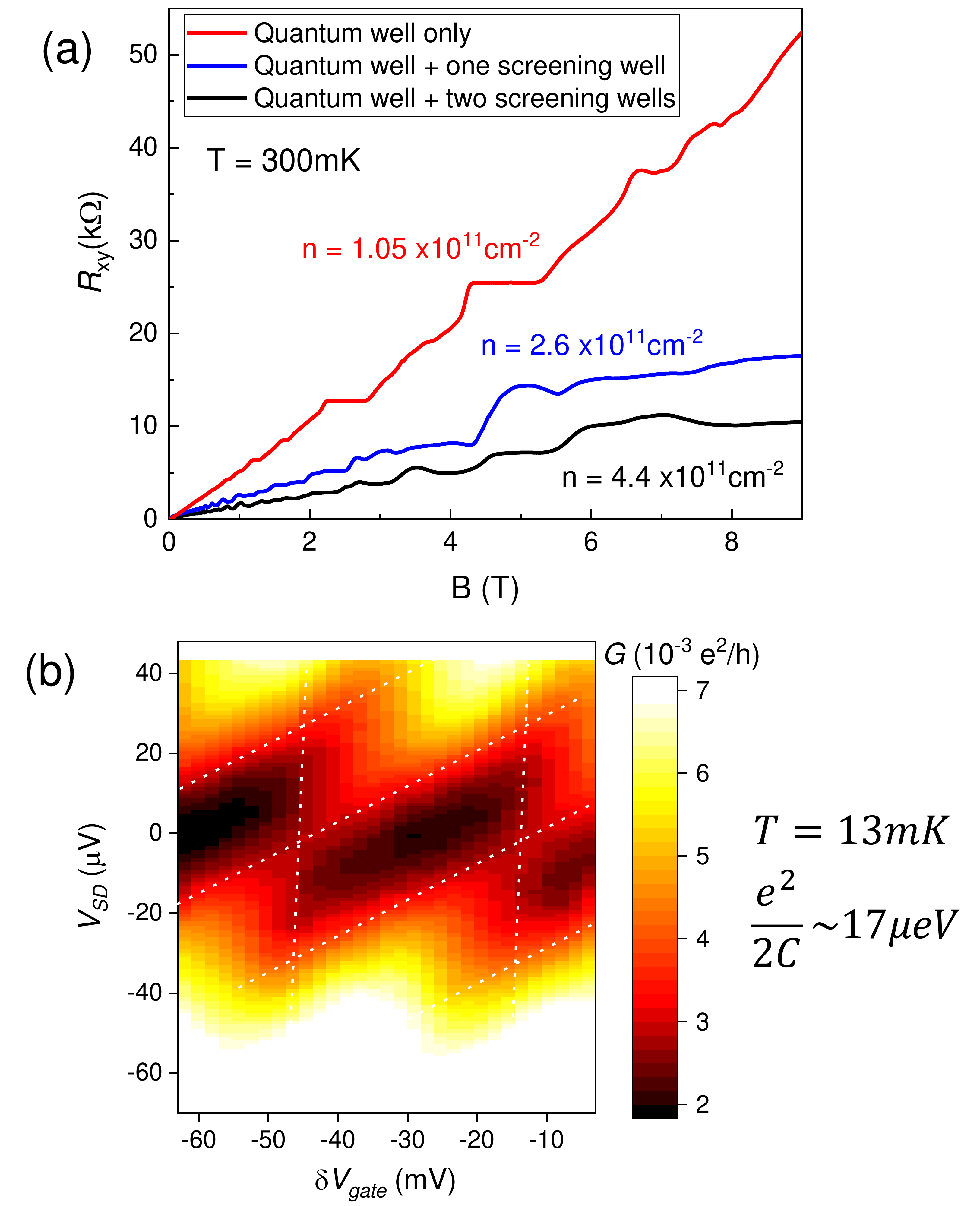}
 \caption{\label{Transport}  (a) Bulk Hall conductance $R_{xy}$ with the top and bottom gates around the contacts grounded (black trace), with -0.29V on the top gate to disconnect the top screening well from the contacts (blue), and with -0.29V on the top gate and -150V on the back gate around the contacts in order to disconnect both screening wells from the contacts so that transport is only measured through the primary quantum well (red). (b) Coulomb blockade measurement at zero magnetic field measured in a dilution refrigerator at base temperature $T=13mK$ showing the differential conductance $\frac{\partial I}{\partial V}$ versus gate voltage and source-drain voltage $V_{SD}$ for the device at zero field showing Coulomb blockade diamonds with charging energy $\frac{e^2}{2C} \sim 17 \mu eV$.}
\end{figure}

Isolation of the screening wells is tested prior to energizing the interferometer gates. In Fig \ref{Transport} (a) we show the evolution of the Hall resistance $R_{xy}$ as the gates that overlay the Ohmics are biased. In the black trace, no bias is applied to the gates, so current flows through all three quantum wells. In the blue trace, the top surface gate around the Ohmic contacts (orange gate in Fig. \ref{Structure} (c)) is negatively biased to deplete the top screening well; the Ohmic contacts are disconnected from the top screening well and transport is only measured through the primary quantum well and bottom screening well. In the red trace, the bottom gate around the Ohmic contacts (red gate in Fig. \ref{Structure} (c)) is also negatively biased to deplete the bottom screening well so that only the primary quantum well is probed; in this case $R_{xy}$ exhibits a much steeper slope and shows clear quantum Hall plateaus and concomittant zeroes in longitudinal resistance (not shown), demonstrating that parallel conduction through the screening wells has been eliminated.

The presence of the screening wells acts to reduce the Coulomb charging energy, characterized by measuring Coulomb blockade through the device at zero magnetic field \cite{Beenakker1991}. Coulomb blockade diamonds (obtained by measuring the differential conductance $\frac{\partial I}{\partial V}$ versus side gate voltage $V_{gate}$ and source drain voltage $V_{SD}$), shown in Fig \ref{Transport} (b),  yield a charging energy $\frac{e^2}{2C} \approx 17\mu eV$. The Coulomb blockade charging energy characterizes the incremental increase of electrostatic energy when an electron is added in the presence all of the other electrons localized in the interior of the device; therefore, this energy may be loosely identified with the bulk-edge coupling constant $K_{IL}$ in Ref.\cite{Halperin2011},  which determines whether the device is in the Coulomb-dominated or Aharonov-Bohm regime. A similarly sized device \textit{without} screening wells would have charging energy $\frac{e^2}{2C} \sim \frac{e^2}{\epsilon r} \approx 200 \mu eV$ (where $r$ is the radius of the dot), indicating that the screening wells are very effective at reducing Coulomb effects.

\section{$\nu = 1$ interference}

\begin{figure*}
\def\ffile{Interference.pdf}
\centering
\includegraphics[width=\linewidth]{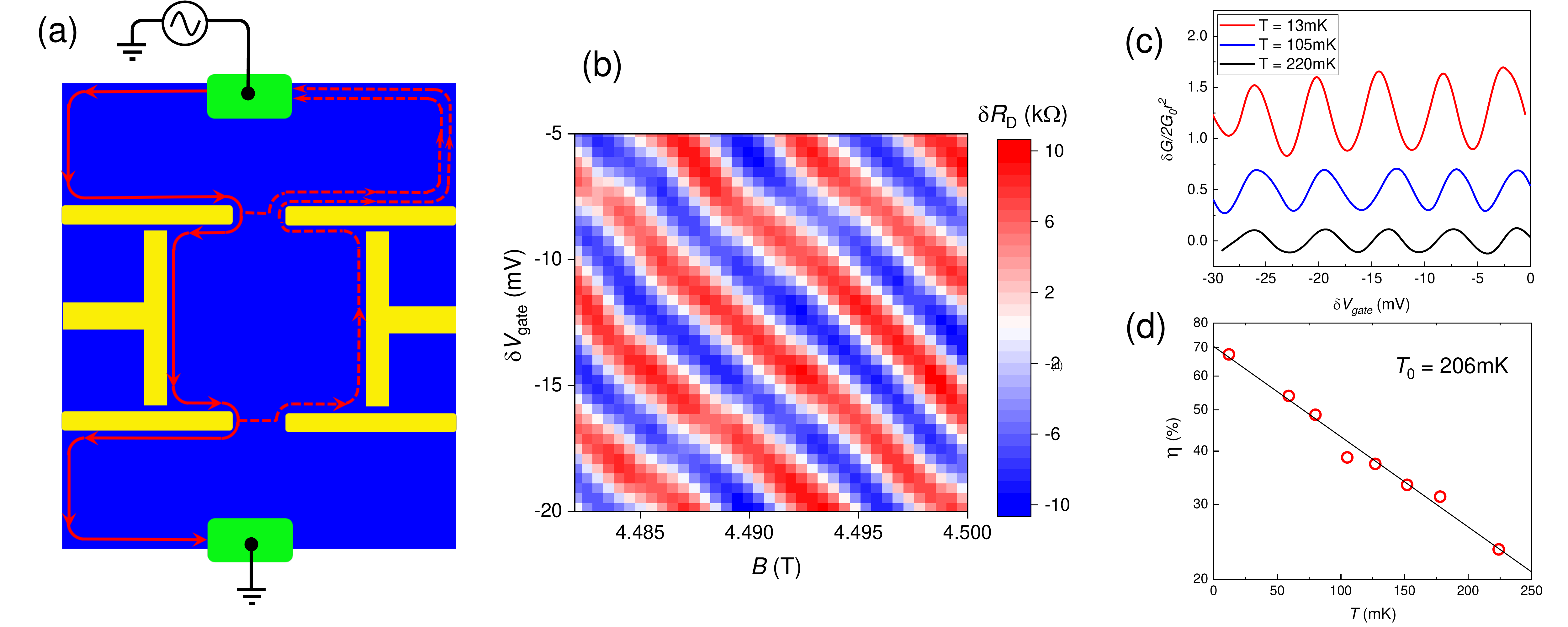}
\caption{\label{Interference} (a) Schematic showing the interference path defined by the interferometer gates at $\nu=1$. (b) Resistance oscillations as a function of magnetic field $B$ and side gate voltage $\delta V_{gate}$ (relative to -1.4V) showing clear Aharonov-Bohm interference. (c) Oscillations in conductance through the device, $\delta G$, divided by the QPC backscattering amplitude, $r^2$, at 13mK (red), 105mK (blue), and 220mK (black). For these measurements each QPC is tuned to approximately 97\% transmission and 3\% reflection ($r^2$ = 0.03). The amplitude of the oscillations clearly decreases as temperature is increased. (d) Coherence factor $\eta$ versus temperature; $\eta$ shows an approximately exponential dependence on temperature with a characteristic decay scale of 206mK. }
\end{figure*}

Next, we operate the device at filling factor $\nu=1$ in the integer quantum Hall regime, where the bulk of the 2DES is insulating and current is carried by chiral edge states. The interference path is shown schematically in Fig. \ref{Interference} (a). Electrons incident from the source contact are backscattered by the two quantum point contacts to the opposite edge, and the two backscattered paths interfere; this is shown schematically in Fig. \ref{Interference} (a). The quantum mechanical Aharonov-Bohm phase accumulated by an electron traversing the interference path is given by $\theta = 2\pi \frac{A_I B}{\Phi_0}$, where $A_I$ is the area of the interference path, B is the magnetic field, and $\Phi _0 \equiv \frac{h}{e}$ is the magnetic flux quantum. The device may be operated by changing the magnetic field B, or by changing the voltage on the side gates to change $A_I$.

At $\nu=1$ the interferometer exhibits strong conductance oscillations, probed by measuring the diagonal resistance $R_D$ across the device. $R_D$ as a function of gate voltage and magnetic field is plotted in Fig. \ref{Interference} (b); the lines of constant phase exhibit negative slope, consistent with the device being in the Aharonov-Bohm regime despite its small size. The magnetic field oscillation period $\Delta B = 5.7mT$, which gives an area of the interference path $A_I = \Delta B/\phi_0 \approx 0.73 \mu m^2$. This area is smaller than the lithographic area of the device, indicating that the 2DES is depleted in a region approximately 180nm wide around the gates; this agrees with simulations of the 2DES density at the edge of the gate (see Supp. Fig. 1). The magnetic field period does not vary significantly with filling factor, consistent with a device in the Aharonov-Bohm regime \cite{Zhang2009, Halperin2007, Halperin2011}. Previous Fabry-Perot interferometry experiments utilizing conventional heterostructures have required a device area of $~20\mu m^2$ in order for Coulomb effects to be small enough for the device to be in the Aharonov-Bohm regime \cite{Zhang2009, Heiblum2009}; unambiguous observation of the Aharonov-Bohm regime in a much smaller device demonstrates the effectiveness of the device design employed here.

For weak backscattering by symmetrically tuned QPCs, conductance oscillations  due to interference obey $\delta G/G_0 = 1-2r^2 \eta \cos(2\pi \frac{AB}{\phi_0})$, where $G_0 \equiv \frac{e^2}{h}$ is the conductance quantum, $r^2$ is the reflection amplitude of the QPCs, and $\eta$ is the coherence factor. We characterize coherence of the interference at $\nu = 1$ by measuring conductance oscillations at different temperatures, plotted in Fig. \ref{Interference} (c); we normalize by dividing by the conductance oscillations $\delta G$ by the reflection amplitude $r^2$, with each QPC tuned to approximately 97\% transmission and 3\% reflection. The coherence factor $\eta$ (defined as the amplitude of $\frac{\delta G}{2G_0 r^2}$)  decays with temperature following an approximately exponential trend, shown in Fig. \ref{Interference} (d), with a characteristic temperature $T_0 = 206$mK. For comparison, in measurements of a Fabry-Perot interferometer in \cite{McClureThesis} $T_0$ was found to be $< 20mK$ for magnetic fields exceeding 1.5T; in measurements of Mach-Zehnder interferometers the largest $T_0$ measured was ~40mK \cite{Roulleau2008}, with larger devices exhibiting smaller $T_0$. The significantly larger $T_0$ observed in our experiment indicates that the smaller size achieved in our device is beneficial to achieving quantum coherence.                 

\section{Edge mode velocity}

\begin{figure*}
\def\ffile{Velocityv2.pdf}
\centering
\includegraphics[width=\linewidth]{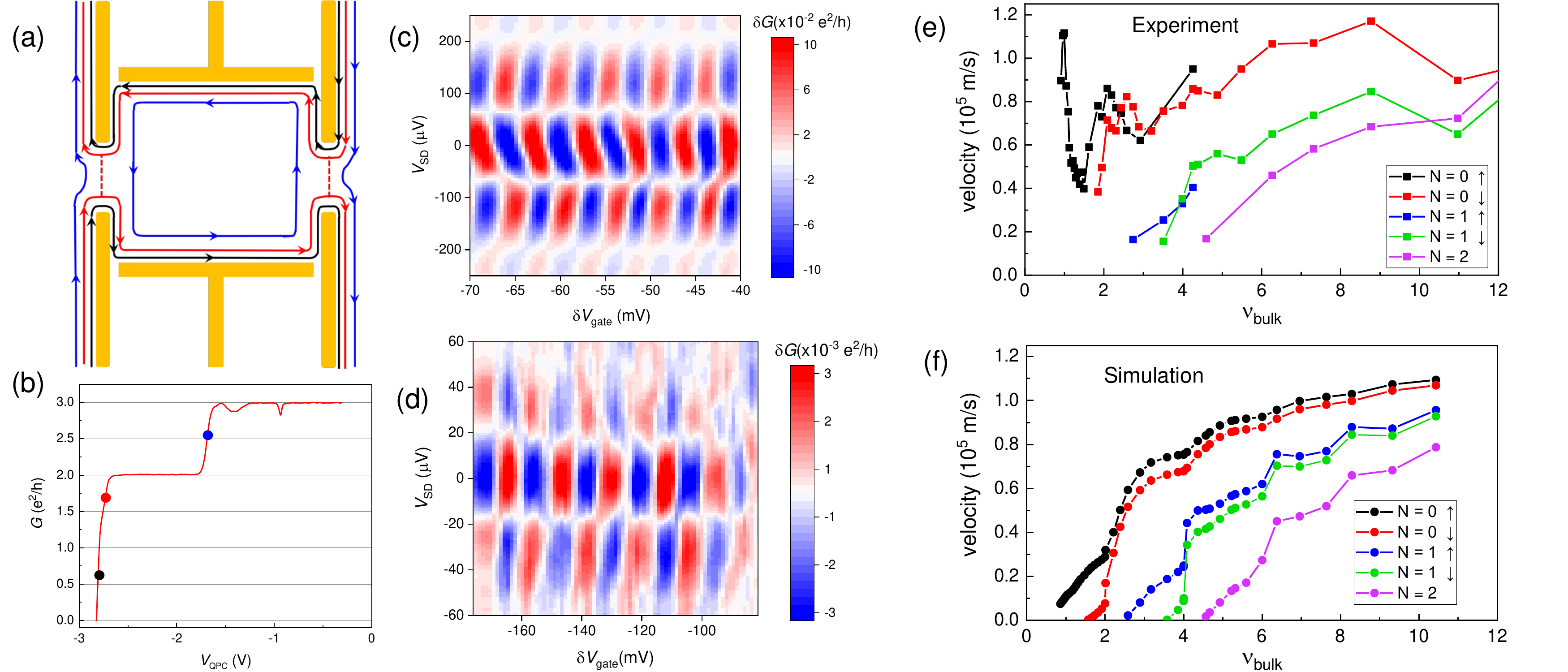}
\caption{\label{Velocity} (a) Schematic showing an interference path with multiple edge states in which the outermost mode is fully transmitted, the innermost mode is fully backscattered by both QPCs, and the middle mode is partially transmitted by both QPCs; in this configuration only the middle mode is interfered. (b) Conductance versus gate voltage for one QPC at B = 1.64T and $\nu_{bulk}=3$ with other interferometer gates grounded. The blue, red, and black circles indicate the operating point for interference of the modes associated with $\nu=3$, $\nu=2$, and $\nu=1$ respectively. (c) Differential conductance at $\nu_{bulk} = 1$ interfering the $\nu=1$ mode and (d) at $\nu_{bulk} = 3$ interfering the innermost $\nu = 3$  mode as a function of side gate voltage and source drain voltage. (e) Edge state velocity extracted from the differential conductance oscillations for different edge modes as a function of bulk filling factor. At magnetic fields below approximately 1.2T (bulk filling factor $\nu=4$) conductance through the QPCs is no longer spin-resolved, so only a single line is displayed for each Landau level. (e) Numerically calculated edge state velocities for the N = 0, 1, and 2 Landau levels. }
\end{figure*}

When the device is operated at lower magnetic field (higher filling factor), multiple integer edge modes are present. In our device it is possible to selectively interfere a particular edge mode by tuning the QPC voltages to partially backscatter that edge, while fully transmitting the outer edges so that only the partially backscattered edge interferes; this is shown schematically in Fig. \ref{Velocity} (a) for the case of bulk filling factor $\nu _{bulk} = 3$, and a corresponding trace of the QPC conductance versus gate voltage is shown in Fig. \ref{Velocity} (b) with the operating points corresponding to the selective interference of each edge state indicated with colored circles.

The interference phase may be additionally modulated by changing the energy $\epsilon$ of injected electrons, which changes the wave-vector $k$. This introduces a phase shift $\delta \theta = \delta \epsilon \frac{\partial k}{\partial \epsilon}L = \frac{\delta \epsilon L}{\hbar v_{edge}} $, where L is the path length around the interference loop and $v_{edge} \equiv \frac{1}{\hbar} \frac{\partial \epsilon}{\partial k}$ is the velocity of the edge state \cite{Wen1997}. $\epsilon$ may be modulated by applying a finite source-drain bias $V_{SD}$ across the device; this results in oscillations in differential conductance as a function of both $V_{SD}$ and flux: $\delta G \propto \cos (2 \pi \frac{AB}{\Phi _0}) \cos (\frac{eV_{SD}L}{2\hbar v_{edge}})$ \cite{McClure2009}. This results in nodes in a ``checkerboard'' pattern when $\delta G$ is measured in the $V_{SD}$ - $V_{gate}$ plane (plotted at $\nu _{bulk} = 1$ in Fig. \ref{Velocity} (c) and for the inner N = 1 mode at $\nu _{bulk} = 3$ in Fig. \ref{Velocity} (d)), with nodes in the interference pattern occurring at $V_{SD} = \pm \frac{ \pi \hbar v_{edge}}{eL}$. The velocity may thus be extracted: $v_{edge} = \frac{eL \Delta V_{SD}}{2\pi \hbar}$ \cite{Heiblum2016}, where $\Delta V_{SD}$ is the spacing between nodes, and we estimate $L$ from the interference area, $L \approx 4 \sqrt{A_I}$. 
 
In Ref. \cite{McClure2009} this method was used to measure edge velocity versus filling factor, but without controlling which edge mode was being interfered; in \cite{Heiblum2016} edge velocity for only the $N = 0$ LL was reported. To our knowledge, measurement of edge velocity for different Landau levels as a function of filling factor has not been demonstrated previously. In Fig. \ref{Velocity} (e) we plot the edge state velocity for the N = 0, N = 1, and N = 2 Landau level edge modes versus bulk filling factor $\nu _{bulk}$. The inner, higher index Landau levels generally have lower velocity and correspondingly lower coherence. At magnetic fields below approximately 1.2T ($\nu _{bulk} = 4$), the QPCs show spin-degenerate conductance plateaus, even though the bulk transport exhibits spin-split quantum Hall states down to ~0.2T. This suggests that although distinct edge states exist, below 1.2T they are too close to one another to be interfered independently; therefore at filling factors $v_{bulk} > 4$ we show a single velocity measurement for each Landau level, while at lower fillings we show both spins when resolved. We also mention that we observe the same period-halving phenomenon in our device that was reported in previous interferometry experiments \cite{Heiblum2015,Heiblum2018}; see Supp. Note 4 and Supp. Fig. 5.  

Much of the magnetic field dependence in Fig. \ref{Velocity} (e) can be understood from the fact that edge currents in the quantum Hall regime are generated by Hall drift: $\vec{v}_{Hall} = \frac{ \vec{E} \times {\vec{B}}}{B^2}$, where $\vec{E}$ is the in-plane electric field at the edge due to the confining potential and $\vec{B}$ is the perpendicular magnetic field. This implies that the edge velocity should increase with decreasing magnetic field (increasing filling factor), and this is indeed the predominant trend observed at filling factors $9<\nu _{bulk}<2$. On the other hand, it must also be considered that the electric field experienced by each edge state also depends on both magnetic field and Landau level index. It can be seen from Fig. \ref{Velocity} (e) that the outer, lower index Landau levels generally have higher edge velocity than the inner, higher index ones. This behavior can be understood from the works of Chklovskii et al. \cite{Chklovskii1992, Chklovskii1993}, who found that the confining potential is steepest at the outer edge, resulting in a higher electric field and thus higher velocity for the outer Landau level edge modes and a smaller electric field and lower velocity for the inner ones.

Numerical simulations of edge transport in the integer quantum Hall regime for the heterostructure used in these experiments have been performed, and are plotted in Fig. \ref{Velocity} (f); see Supplementary Note 1 and Ref. \cite{Harshad2018} for an in-depth review. In these simulations, the spatially varying in-plane electric field is self-consistently evaluated for the Landau level density of states, considering the electrostatic effects of the heterostructure, doping, surface states and gates. The velocity is obtained by solving quantum transport (non-equilibrium Green's function) equations at the Fermi level.

The simulations show good qualitative and quantitative agreement with the experimental results over the range of filling factor $2<\nu _{bulk} <10$. At lower filling $\nu _{bulk}<2$, the edge velocity exhibits non-monotonic behavior, which may be due to the impact of electron-electron interactions which become increasingly important at high magnetic field. Non-monotonic behavior at low filling was also reported in Ref. \cite{Heiblum2016}. Our simulations employ a mean-field Hartree approximation that does not capture many-body effects.

Additionally, the edge velocities also exhibit non-monotonic behavior at high filling $\nu _{bulk} > 10$. A possible explanation for this is that at low fields when the magnetic length becomes comparable to the length scale of the confining potential at the edge, charge transport may occur via skipping orbits, resulting in different behavior than observed at higher fields \cite{Montambaux2011, McClure2009}. It is reasonable for this to occur at  $~ \nu _{bulk} = 10$; here the magnetic length is $~$39nm, and simulations indicate that the length scale of the confining potential is approximately 40nm (see supplementary Fig. 1). This effect is not captured in the simulations as the magnetic length approaches the Debye length.

\section*{Fractional quantum Hall regime}

We turn now to results in the fractional quantum Hall regime. In previous experiments with small Fabry-Perot devices Coulomb-dominated or Coulomb blockade oscillations have been observed in fractional states \cite{McClure2012, Heiblum2010, Goldman2005, Goldman2007}.  Willet et al.\cite{Willett2009, Willett2013} reported oscillations at $\nu=5/2$ consistent with Aharonov-Bohm interference of charge $e/4$ and $e/2$ excitations. However, oscillations with negatively sloped lines of constant phase in the gate voltage-magnetic field plane (a sine qua non of Aharonov-Bohm regime interference) have not been previously reported. Edge states in the fractional quantum Hall regime are predicted to have remarkably different properties from those in the integer states; in particular, the current-carrying quasiparticles may carry fractional charge. In the fractional case, the Aharanov-Bohm interference phase is modified \cite{Halperin2011}:  

\begin{equation} \label{FractAB}
\theta = 2\pi \frac{e^*}{e} \frac{A_I  B}{\Phi _0}
\end{equation}

Eqn. \ref{FractAB} indicates that quasiparticle charge may be extracted from gate voltage oscillation periods according to the relationship $\frac{e^*} {e} = \frac{\Phi_0}{B \Delta V_{gate}\frac{\partial A_I}{\partial V_{gate}}}$, where $\Delta V_{gate}$ is the gate voltage oscillation period and $\frac{\partial A_I}{\partial V_{gate}}$ is the lever arm relating change in gate voltage to the change in interference path area. $\frac{\partial A_I}{\partial V_{gate}}$ may be determined from the gate voltage period at integer states, where the interfering charge is simply $e$; a linear fit of $\Delta V_{gate}$ versus $1/B$ yields $\frac{\partial A_I}{\partial V_{gate}} = 1.8\times 10^{-13} m^2V^{-1}$ (gate and magnetic field periods are shown in Supp. Fig. 3). We mention that interference at $\nu_{bulk} = 1/3$ was found to be reproducible using a range of different gate voltages as well as after thermal cycling the device to room temperature; see Supp. Note 3 and Supp. Fig. 4.

In both the Laughlin \cite{Laughlin1983} and composite fermion \cite{Jain1989, JainBook} theories the $\nu = 1/3$ FQHE state is predicted to support quasiparticles with charge $e^* = e/3$. At $\nu = 1/3$ ($B = 13T$), we observe conductance oscillations as a function of gate voltage and magnetic field similar to those at integer states; the oscillations have gate voltage period $\Delta V_{gate}=6.1mV$; this yields an interfering quasiparticle charge $e^* = e\frac{\Phi_0}{B \Delta V_{gate}\frac{\partial A_I}{\partial V_{gate}}} = 0.29e$, in good agreement with the theoretical predictions. This supports previous experimental results utilizing shot noise \cite{Heiblum1997}, resonant tunneling \cite{Goldman1995, Goldman1995Science}, and Coulomb blockade \cite{McClure2012}. 

Next we discuss the $\nu = 2/3$ FQHE state, which is the hole-conjugate state to $\nu = 1/3$ \cite{Girvin1984}. Several edge structures have been proposed for the $\nu = 2/3$ state. Motivated by a picture in which the $\nu = 2/3$ consists of a $\nu = 1/3$ hole state imposed upon a $\nu = 1$ background, MacDonald proposed that the $\nu = 2/3$ edge should consist of an inner edge mode of charge $e^* = -e/3$ and an outer edge with $e^* = e$ \cite{Macdonald1990}. Chang \cite{Chang1990} and Beenakker \cite{Beenakker1990} constructed models consisting of two $e^* = e/3$ edge modes; a later work indicated that a transition from the MacDonald edge structure to the Chang-Beenakker edge structure should occur as the confining potential is tuned from sharp confinement to soft confinement \cite{Meir1993}. Yet another edge model was proposed by Kane, Fisher, and Polchinski in which the presence of disorder leads to a single $e^* = 2e/3$ charged edge mode and a counterpropogating neutral mode \cite{Kane1994}.

We measure conductance oscillations at $\nu = 2/3$ ($B = 6.8T$) with $\Delta V_{gate}=3.7mV$, yielding a quasiparticle charge $e^* = e\frac{\Phi_0}{B \Delta V_{gate}\frac{\partial A_I}{\partial V_{gate}}} = 0.93e$, which suggests interference of an integrally charged edge mode. Presence of an integrally charged mode suggests that the Macdonald edge structure holds in our device. However, we do not find evidence for interference of a fractionally charged $e^* = -e/3$  mode at $\nu = 2/3$, even if the QPC bias is tuned to reduce backscattering. A possible explanation for this is that $e^* = -e/3$ should have a significantly smaller velocity due to being an inner mode; therefore, it will have lower phase coherence, making it very difficult to observe. Smaller device size or lower experimental temperatures might make measurement of the $-e/3$ mode possible.

It is noteworthy that our observation of an integrally charged mode differs from previous experimental findings, in which shot noise and Coulomb blockade measurements suggested a different edge structure consisting of two $e^* = e/3$ charge modes and two neutral modes \cite{Heiblum2009, Heiblum2017}, with no integrally charged mode observed. A possible explanation for this discrepancy is that our sample may have a sharper confining potential due to the short setback of the screening wells (see Supp. Note 2 and Supp. Fig. 2), resulting in our device supporting the edge structure described in Ref.\cite{Macdonald1990}. We mention that a sharp confining potential may also be beneficial for measuring interference at the $\nu = 1/3$ state by preventing edge reconstruction and the proliferation of neutral edge modes \cite{Wan2002,Joglekar2003,Hu2009} which may cause dephasing \cite{Gefen2016,Gefen2015}; neutral modes have been detected at $\nu = 1/3$ and numerous other fractional quantum Hall states in standard GaAs structures without screening wells \cite{Heiblum2014}. 

Finally, we remark that although we have observed Aharonov-Bohm interference of fractionally charged quasiparticles at the $\nu = 1/3$ fractional quantum Hall state, we have not observed the fractional braiding statistics predicted for these quasiparticles \cite{Halperin1984,JainBook}. It has been suggested that increasing the flux through the interferometer by one flux quantum should result in the addition of one quasiparticle into the area of the device in order to keep the system charge neutral; this should result in an interference phase jump $\Delta \theta _{anyon} = 4\pi /3$ at the $\nu = 1/3$ state\citep{Wen1997, Halperin2011}. We appear to measure only the Aharonov-Bohm phase when magnetic field is varied, suggesting that adding flux does not introduce quasiparticles in our device. Critically, the $\nu = 1/3$ state has a large energy gap for the creation of quasiparticles measured to be $\sim 700 \mu eV$ in a 2DES of similar density \cite{Stormer1993}. This energy is more than an order of magnitude larger than the measured charging energy in our device ($\frac{e^2}{2C} \sim 17 \mu V$), which suggests that when magnetic field is varied it may be energetically favorable for the primary quantum well to remain at fixed filling factor (without creating quasiparticles) rather than fixed sheet density, with the energy cost of the variations in quantum well density reduced by the screening wells. When the experiment is performed at fixed filling factor it is expected that only the Aharonov-Bohm  phase of the quasiparticles will be observed when magnetic field and side gate voltage are varied \cite{Wen1997,Halperin2006}, consistent with  our observations. An alternative method to introduce quasiparticles and measure braiding statistics would be to directly manipulate the electrostatic potential with a gate in the center of the interferometer \cite{Wen1997,Halperin2006}; efforts are underway to fabricate devices with this type of gate.

\section*{Conclusions}
We have demonstrated a small electronic Fabry-Perot interferometer in which Coulomb effects are minimized, facilitating measurement of highly coherent Aharonov-Bohm interference of both integer and fractional quantum Hall edge modes. Selective population of inner and outer edge states in the integer quantum Hall regime allow determination of the velocity of each mode. Measurement of Aharonov-Bohm interference at the $\nu = 1/3$ and $\nu = 2/3$ fractional quantum Hall states paves the way towards direct observation of fractional braiding statistics with modest modifications to device design.  

\begin{figure}
\def\ffile{Fractionsv2.pdf}
\centering
\includegraphics[width=\linewidth]{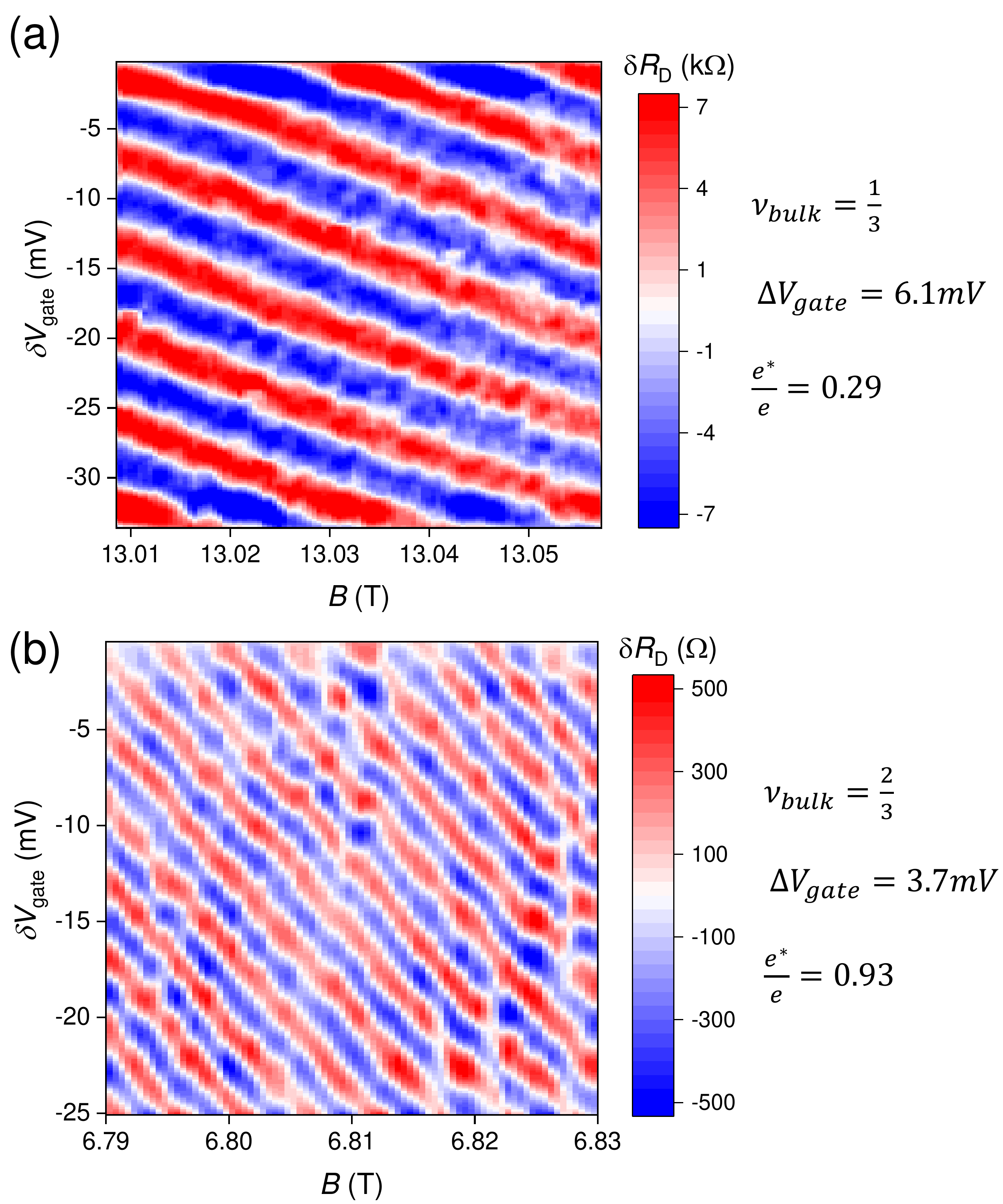}
\caption{\label{Fractions} (a) Aharonov-Bohm conductance oscillations at $\nu=1/3$. (b) Aharonov-Bohm conductance oscillations at $\nu=2/3$.}
\end{figure}

\section*{Methods}
The primary quantum well was measured to have bulk electron density $n = 1.05 \times 10^{11} cm^{-2}$ and mobility $\mu = 7 \times 10^6 cm^2 V^{-1} s^{-1}$ measured after full device fabrication and in the dark.

The device was fabricated by: (1) optical lithography and wet etching to define the mesa; (2) deposition of In/Sn Ohmic contacts; (3) electron beam lithography and electron beam evaporation (10nm Ti/15nm Au) to define the interferometer gates; (4) optical lithography and electron beam evaporation (20nm Ti/150nm Au) to define the bondpads and the surface gates around the Ohmic contacts; (5) mechanical polishing to thin the GaAs substrate; (6) optical lithography and electron beam evaporation (200nm Ti/150nm Au) to define the backgates.

The device was measured in a dilution refrigerator with  base mixing chamber temperature $T=13mK$. Extensive heat sinking and filtering are used to achieve low electron temperatures. Standard 4-terminal and 2-terminal lock-in amplifier techniques were used to probe the diagonal resistance and conductance across the device. 

\subsection*{Data Availability}
The data that supports the plots within this paper and other findings of this study are available from the corresponding author upon reasonable request.

\section*{Acknowledgments} This work was supported by the Department of Energy, Office of Basic Energy Sciences, under Award number DE-SC0006671.  Additional support for sample growth from the W. M. Keck Foundation and Nokia Bell Labs is gratefully acknowledged.

\section*{Author Contributions} J.N. and M.M. designed the heterostructures and experiments. S.F., S.L. and G.G. conducted molecular beam epitaxy growth.  J.N. fabricated the devices, performed the measurements, and analyzed the data with input from M.M. H.S. and R.R. performed numerical simulations. J.N. and M.M wrote the manuscript with input from all authors.

\section*{Competing financial interests}
The authors declare no competing financial interests.

\end{document}

\title{Supplementary Information for ``Aharonov-Bohm interference of fractional quantum Hall edge modes''}
\renewcommand{\figurename}{SUPP. FIG.}

\maketitle
\section{Supplementary Note 1: Simulation details}

The velocities of edge states in the integer quantum Hall regime are calculated using quantum transport simulations. These quantum transport simulations require an electrostatic potential in the heterostructure including the gates to model the edges of the 2DEG. This is obtained by self-consistently solving effective-mass Schr\"odinger and Poisson equations for each magnetic field and gate voltage values. The equations are discretized on a finite element mesh that captures the heterostructure geometry. This method is described in \cite{Harshad2018}.

The sheet densities in the three quantum wells were experimentally measured as shown in Fig. 3 (a) of the main text. The sheet densities are highly dependent on the donor ionization energy and the screening well widths, which determine the confinement energies. The calculated screening and main well sheet densities were first matched with the experiments by tuning the donor ionization energy and bottom screening well width in quasi 1-D Schr\"odinger-Poisson simulations. These parameters were later used in 3-dimensional electrostatic simulations of the heterostructure.

The charge density on the exposed top surface was calculated by assuming a Schottky barrier of 0.7 eV at the top surface in quasi 1-D Schr\"odinger-Poisson simulations. This charge density was later used in 3-D electrostatic simulations to define a fixed electric field on the top surface for the frozen surface charge model.

\begin{figure*}[p]
\def\ffile{SuppFigEdge.pdf}
\centering
\includegraphics[width=\linewidth]{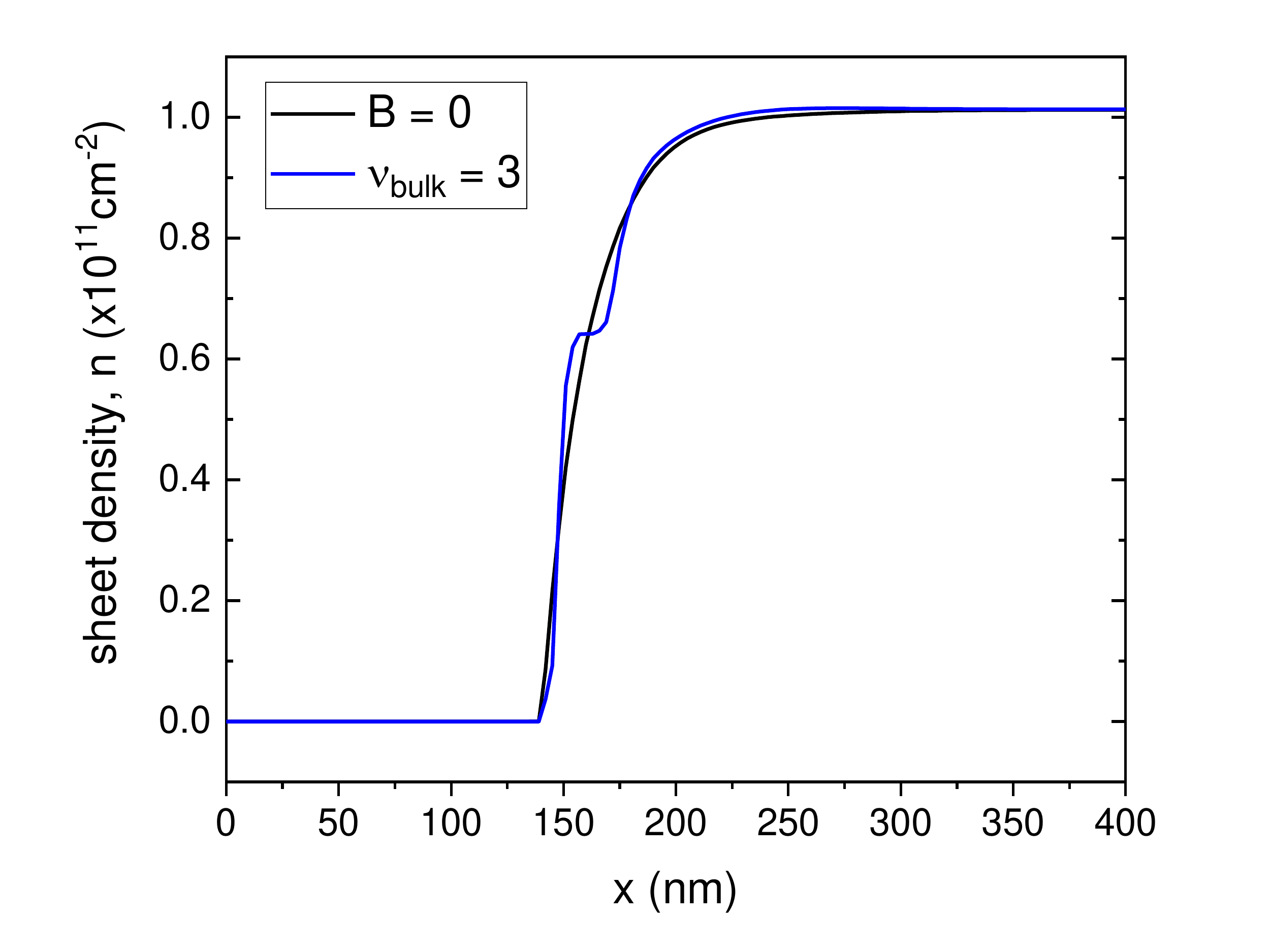}
\caption{\label{Edge} Simulation of the electron sheet density at the edge of the gate versus lateral distance $x$ from the gate at zero magnetic field (black line) and at $\nu_{bulk} = 3$ (blue line). The edge of the gate is located at $x = 0$. For the simulation the gate bias is set at -1.8V to match the experimental side gate bias. The simulation indicates that the 2DES is depleted in a region extending approximately 150nm from the edge of the gate, which is consistent with the experimental finding that the effective area extracted from Aharonov-Bohm oscillations is smaller than the lithographic area.}
\end{figure*}

\maketitle
\section{Supplementary Note 2: Single Interface Structure}
\begin{figure*}[p]
\def\ffile{SuppSingleInterface.pdf}
\centering
\includegraphics[width=\linewidth]{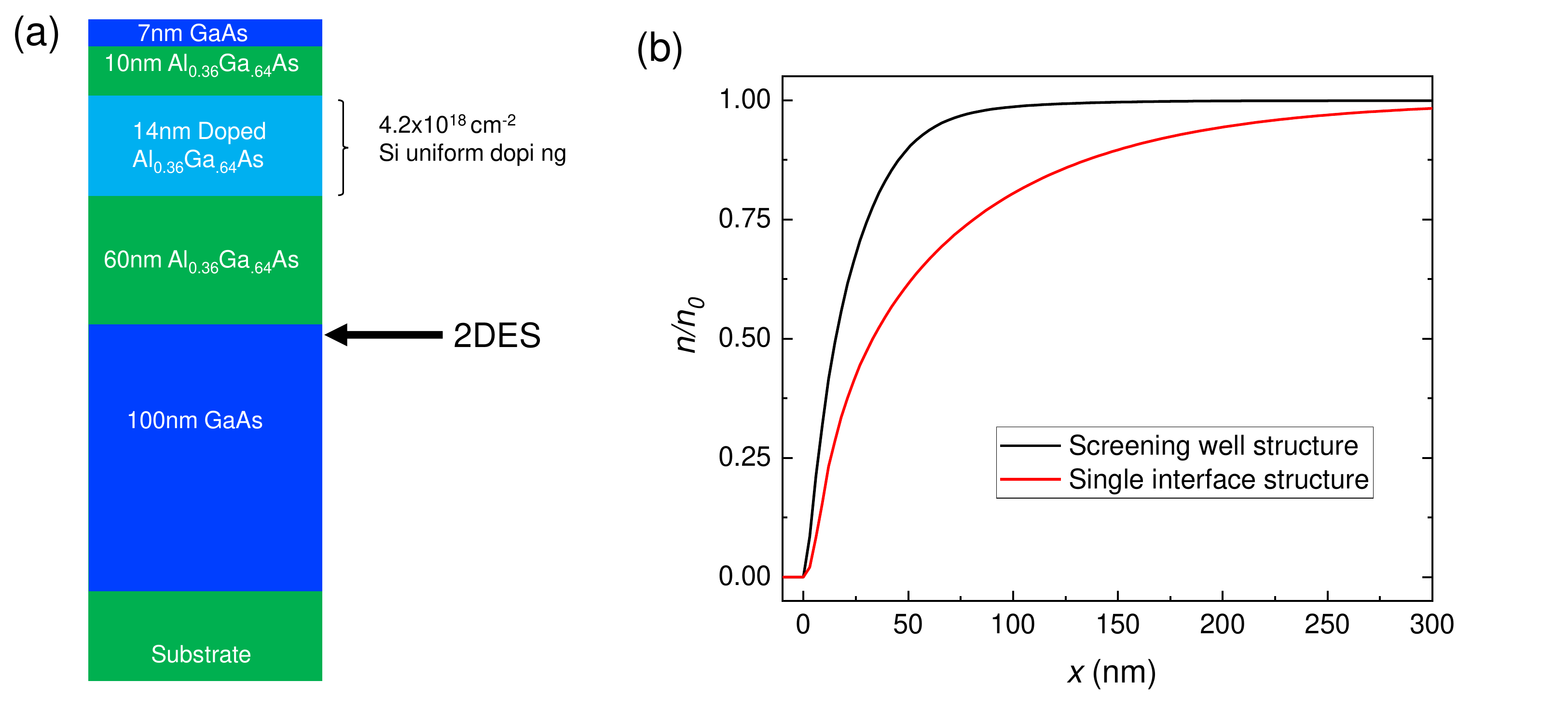}
\caption{\label{SingleInterface} (a) Layer stack for a traditional GaAs/AlGaAs heterostructure without screening wells. The 2DES forms at the interface between AlGaAs and GaAs, 91nm below the surface. This structure has simulated bulk sheet density $n_0 = 1.34\times 10^{11} cm^{-2}$. (b) Numerically simulated density profile of the 2DES edge adjacent to the gate for the screening well structure (black) and the single interface structure (red). The data is normalized  by dividing by the bulk sheet density $n_0$ for each structure. The point $x = 0$ indicates the point on each structure where the 2DES density drops to zero. The screening well structure has a much sharper density profile indicating a sharper confining potential; the length scale over which the density goes from zero to the bulk density is $\sim 3\times$ smaller than for the single interface structure.  }
\end{figure*}

We emphasize that the high density screening well structure we utilize for these experiments is atypical. In Supp. Fig. \ref{SingleInterface} (a) we show a layer stack for a more standard heterostructure without screening wells. In Supp. Fig. \ref{SingleInterface} (b) we show simulations of the electron density at the edge of the gate for the screening well structure and single interface structure (black and red lines respectively). In both cases the applied gate voltage is -1.8V. A key result is that the screening well structure transitions from zero density to the bulk 2DES density over a much shorter length scale than the single interface structure, indicating a much sharper confining potential. The simulation indicates an approximately 3 times sharper edge profile for the screening well structure over the standard heterostructure design.

\begin{figure*}[p]
\def\ffile{SuppFigBothPeriods.pdf}
\centering
\includegraphics[width=\linewidth]{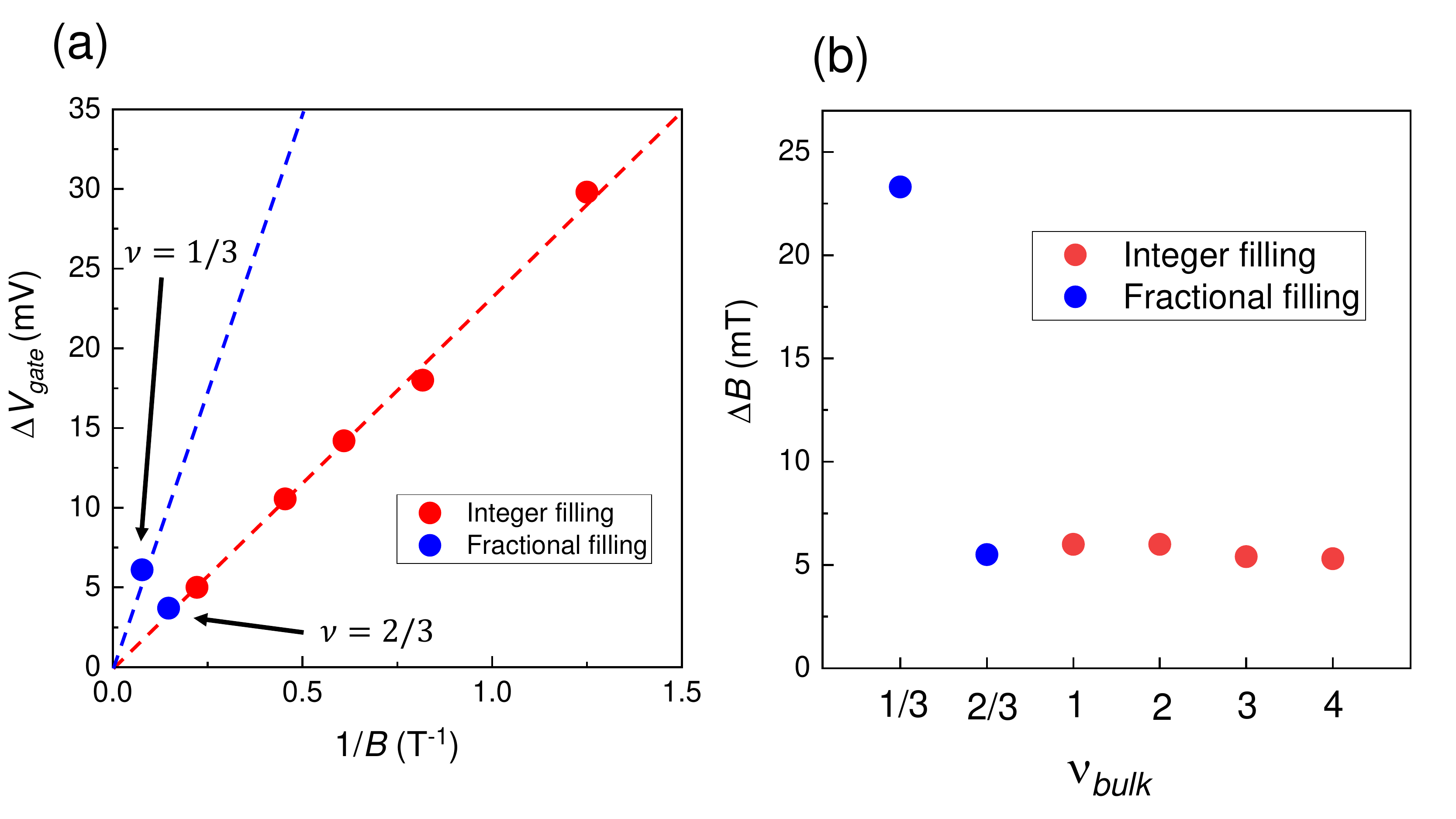}
\caption{\label{GatePeriod} (a) Side gate oscillation period $\Delta V_{gate}$ versus inverse magnetic field at integer quantum Hall states (red circles) and fractional quantum Hall states (blue circles). The dashed red line indicates a linear fit through zero of the integer gate periods, yielding a lever arm  $\frac{\partial A_I}{\partial V_{gate}} = 1.78\times 10^{-13} m^2V^{-1}$. The blue dashed line indicates a line with 3$\times$ larger slope, which would correspond to a quasiparticle charge $e^* = e/3$. The gate period at $\nu_{bulk} = 1/3$ falls close to the blue line, consistent with fractional charge $e^* = e/3$, whereas the period at $\nu_{bulk} = 2/3$ is close to the red line, suggesting integral interfering charge. The integer gate periods used are for the $N =0$ LL; for $\nu_{bulk} \geq 2$ the periods for the spin-down edge state are used to avoid influence of the period-halving phenomenon discussed in Supplementary Note 4. (b) Magnetic field periods $\Delta B$ at different quantum Hall states. Integer states as well as $\nu = 2/3$ have a consistent period of approximately 5.5mT. 
}
\end{figure*}

\maketitle
\section{Supplementary Note 3: Reproducibility}
The Aharonov-Bohm interference observed at the fractional states $\nu_{bulk} = 1/3$  was robust against reasonable variations of QPC voltage and could be reproduced with the QPCs more pinched off. In Supplementary Fig. \ref{Reproduce} (a) we show Aharonov-Bohm oscillations measured with QPCs tuned to the regime of strong backscattering ($G \sim 0.02e^2/h$). The gate voltage and magnetic field oscillation periods are consistent with those presented in the main text, where the device was tuned to weak backscattering. The signal to noise ratio was somewhat worse in this configuration, but the most prominent features of Aharonov-Bohm oscillatory behavior was preserved.

Additionally, the oscillations at $\nu_{bulk} = 1/3$ and $\nu_{bulk} = 2/3$ were found to be repeatable after thermal cycling the device to room temperature and then cooling back to low temperature. In Supplementary Fig. \ref{Reproduce} (b) we show oscillations at $\nu_{bulk} = 1/3$ in a second cooldown of the device. 

\begin{figure*}[p]
\def\ffile{SuppFigReproduce.pdf}
\centering
\includegraphics[width=\linewidth]{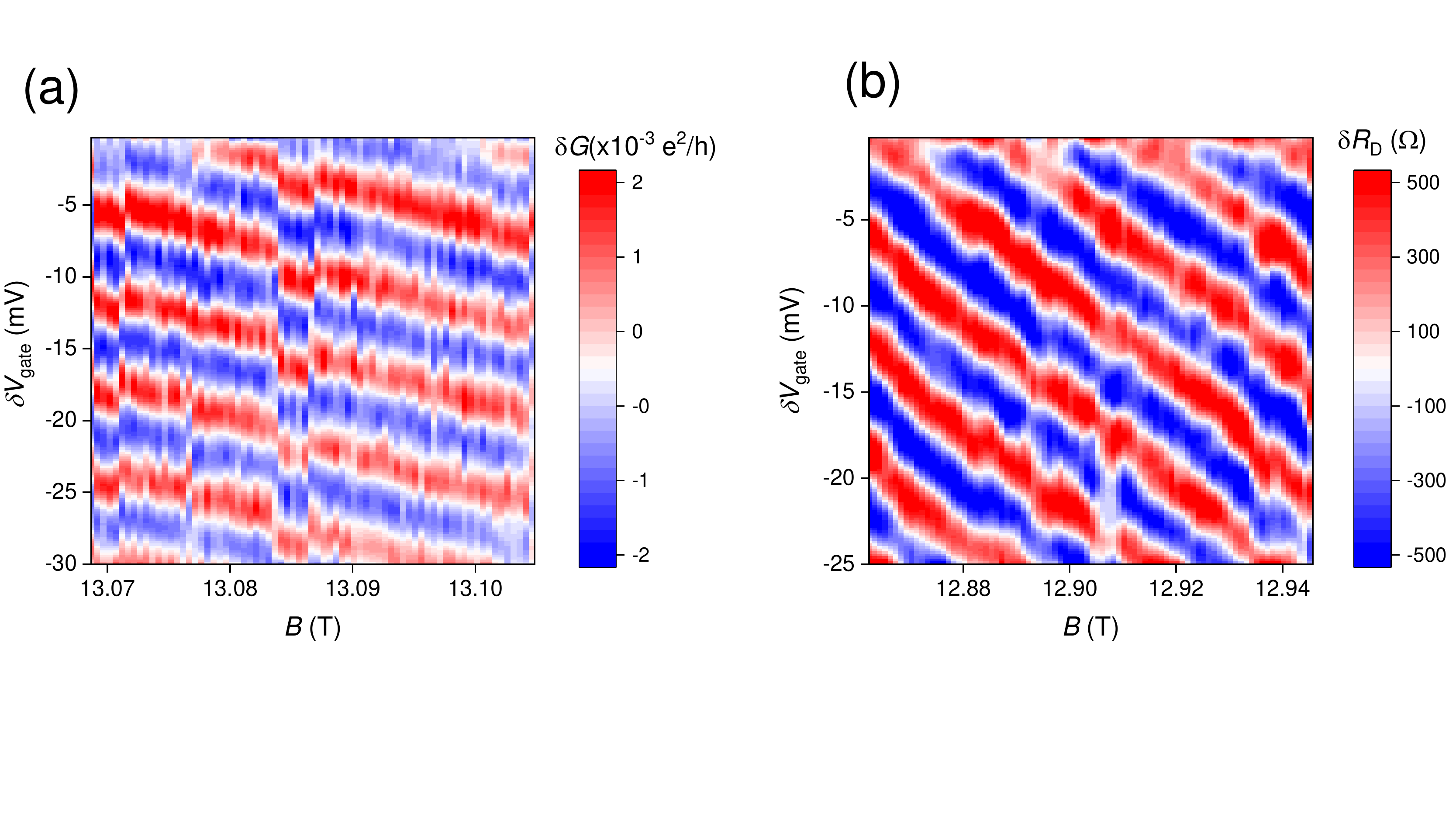}
\caption{\label{Reproduce} (a) Aharonov-Bohm regime oscillations at $\nu _{bulk} = 1/3$ with stronger backscattering. Voltages on the top and bottom QPCs are -1.66V and -2.59V respectiveley; side gate voltages on the vertical axis are relative to -2.1V. The gate voltage and magnetic field periodicities are nearly the same as in the data shown in the main text (b) Oscillations at $\nu _{bulk} = 1/3$ in a second cooldown after cycling the device to room temperature. The fractional state interference was found to be reproducible in this cooldown. }
\end{figure*}

\section{Supplementary Note 4: Period Halving}
\begin{figure*}[p]
\def\ffile{SuppFigPeriodHalving.pdf}
\centering
\includegraphics[width=\linewidth]{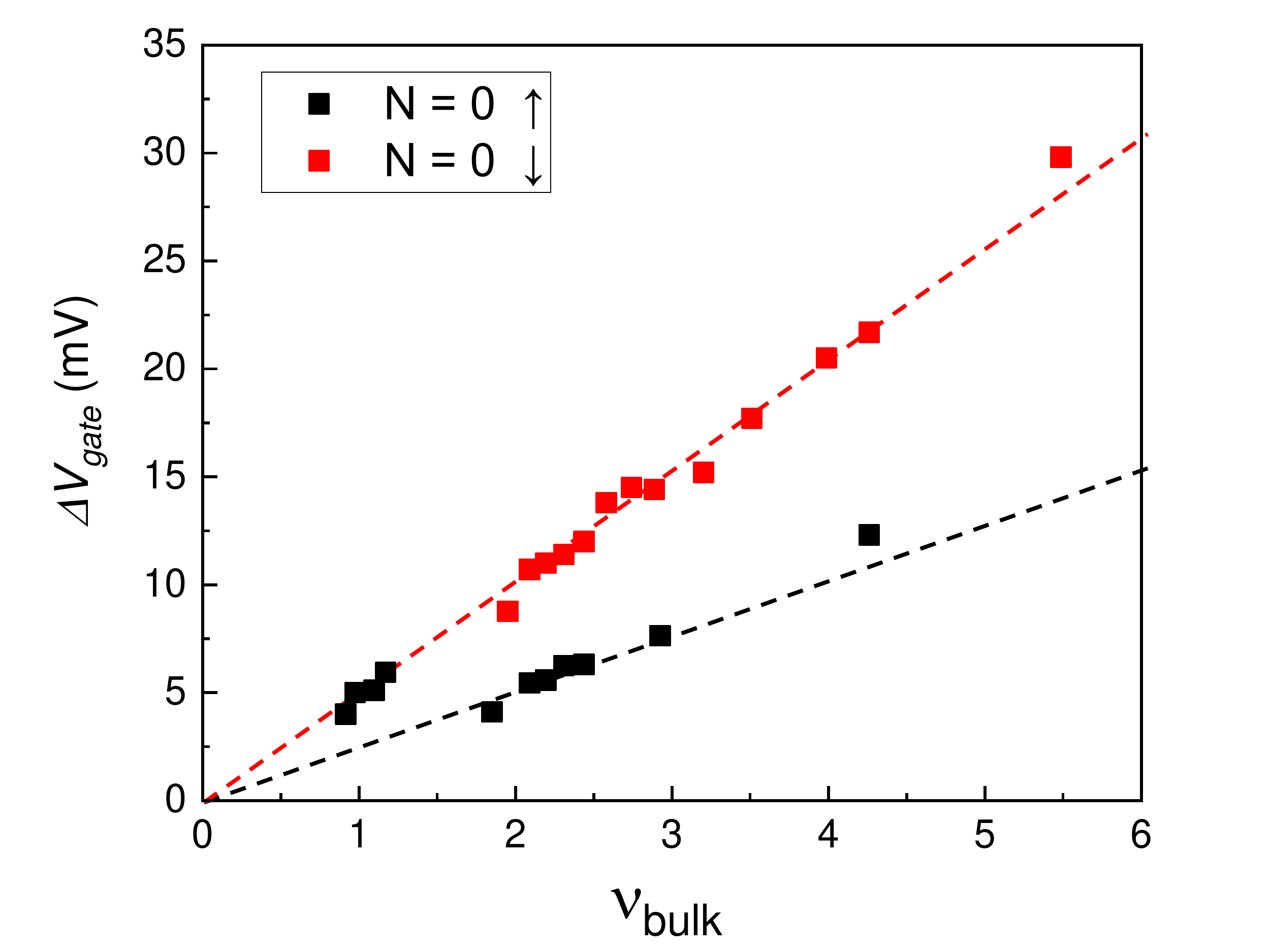}
\caption{\label{PeriodHalving}
Gate voltage period $\Delta V_{gate}$ versus bulk filling factor $\nu _{bulk}$ for the N = 0 Landau level spin-up (black squares) and spin-down (red squares) edge states. The red line indicates a linear fit through zero of spin-up data points for $\nu_{bulk} \leq 1.8$ as well as all the spin-down data points. The black dashed line indicates a linear fit through the spin-up data points for $\nu > 1.8$. The red dashed line has a slope of 5.12 mV, while the black dashed line has a slope of 2.57 mV, which is consistent with N = 0 spin-up LL edge state exhibiting a halved period Aharonov-Bohm behavior when the spin-down edge state is present, consistent with the behavior discussed in Refs. \cite{Heiblum2015,Heiblum2018}.  
}
\end{figure*}

In previous experiments in Fabry-Perot interferomters it was reported that the outermost N = 0, spin up edge mode exhibits half the expected Aharonov-Bohm period in a certain range of filling factor $2.5 \lessapprox \nu _{bulk}\lessapprox 4.5$ \cite{Heiblum2015,Heiblum2018}. We also observe this phenomenon in our device over a similar range of filling factor. The gate voltage periods $\Delta V_{gate}$ of the N = 0 spin up and spin down edge states are plotted versus bulk filling factor $\nu _{bulk}$ in Supp. Fig. \ref{PeriodHalving}. In a simple Aharonov-Bohm picture $\Delta V_{gate}$ would be expected to be proportional to $\nu_{bulk}$; however, for $\nu _{bulk}\gtrapprox 2$, the spin-up edge state has approximately half the expected period, consistent with the period-halving phenomenon reported in the previous experiments. It has been proposed that this occurs due to electron pairing mediated by interactions, resulting in a doubling of the effective interfering charge and thus a halved Aharonov-Bohm period \cite{Heiblum2015,Heiblum2018}.

